\documentclass[twocolumn]{aa}
\usepackage{graphicx}
\usepackage{epstopdf}
\usepackage[intlimits,sumlimits]{amsmath}
\usepackage{deluxetable}
\usepackage{times,epsfig} 
\usepackage{adjustbox}

\usepackage{amssymb}
\usepackage{mathrsfs} 
\usepackage{dirtytalk}
\usepackage[switch]{lineno}
\usepackage{natbib}
\bibpunct{(}{)}{;}{a}{}{,} 
\usepackage{caption}

\usepackage{amsmath}
\usepackage[english]{babel}
\usepackage{longtable}
\usepackage{afterpage}
\usepackage{url}
\usepackage{hyperref}
\usepackage{xcolor}
\definecolor{dark-red}{rgb}{0.4,0.15,0.15}
\definecolor{dark-blue}{rgb}{0.15,0.15,0.4}
\definecolor{medium-blue}{rgb}{0,0,0.5}
\hypersetup{
    colorlinks, linkcolor={dark-red},
    citecolor={dark-blue}, urlcolor={medium-blue}
}

\newcommand{\beqa}{\begin{eqnarray}} 
\newcommand{\eeqa}{\end{eqnarray}}

\newcommand{\bsub}{\begin{subequations}}
\newcommand{\esub}{\end{subequations}}
\newcommand{\beal}{\begin{align}}
\newcommand{\ealn}{\end{align}}

\newcommand{\Msun}{{\ensuremath{\mathrm{M}_{\odot}}}}
\newcommand{\msun}{{\ensuremath{\mathrm{M}_{\odot}}}}

\begin{document}
\title{On the maximum luminosities of normal stripped-envelope supernovae}
\subtitle{- brighter than explosion models allow.}

\author{J. Sollerman\inst{1} 
\and{S. Yang}\inst{1}
\and{D. Perley\inst{2}
\and{S. Schulze}\inst{3}
\and{C. Fremling}\inst{4}
\and{M. Kasliwal}\inst{4}
\and{K.  Shin}\inst{4}
\and{B. Racine}\inst{5}
}
\institute{Department of Astronomy, The Oskar Klein Center, Stockholm University, AlbaNova, 10691 Stockholm, Sweden
\and{Astrophysics Research Institute, Liverpool John Moores University, Liverpool Science Park, 146 Brownlow Hill, Liverpool L35RF, UK} 
\and{Department of Physics, The Oskar Klein Center, Stockholm University, AlbaNova, 10691 Stockholm, Sweden}
\and{Cahill Center for Astrophysics, California Institute of Technology, 1200 E. California Blvd. Pasadena, CA 91125, USA}
\and{Aix Marseille Univ, CNRS/IN2P3, CPPM, Marseille, France}
}}

\date{}
\abstract
    {Stripped-envelope supernovae (SE SNe) of Type Ib and Type Ic are thought to result from explosions of massive stars having lost their outer envelopes. The favoured explosion mechanism is by core-collapse, with the shock later revived by neutrino heating. However, there is an upper limit to the amount of radioactive $^{56}$Ni that such models can accomplish. Recent literature point to a tension between the maximum luminosity from such simulations and observations.}  
    {We use a well characterized sample of SE SNe from the Zwicky Transient Facility (ZTF) Bright Transient Survey (BTS). We scrutinize the observational caveats regarding estimating the maximum luminosity (and thus the amount of ejected radioactive nickel) for the members of this sample.}
    {We employ the strict selection criteria for the BTS to collect a sample of spectroscopically classified normal Type Ibc SNe for which we use the ZTF light curves to determine the maximum luminosity. We cull the sample further based on data quality, light-curve shape, distance and colors, and examine uncertainties that may affect the numbers. The methodology of the sample construction from this BTS sample can be used for many other future investigations.} 
    {We analyze observational data, consisting of optical light curves
     and spectra, for the selected sub-samples. In total we use 129 Type Ib or Type Ic BTS SNe with an initial rough luminosity distribution peaked at M$_r = -17.61 \pm 0.72$, and where 36\% are apparently brighter than the theoretically predicted maximum brightness of M$_r = -17.8 $. When we further cull this sample to ensure that the SNe are normal Type Ibc with good LC data within the Hubble flow, the sample of 94 objects has M$_r = -17.64 \pm 0.54$. A main uncertainty in absolute magnitude determinations for SNe is the host galaxy extinction correction, but the reddened objects only get more luminous after corrections. If we simply exclude objects with red, unusual or uncertain colors, we are left with 14 objects at M$_r = -17.90 \pm 0.73$, whereof a handful are most certainly brighter than the suggested theoretical limit.
     The main result of this study is thus that normal SNe Ibc do indeed reach luminosities above 10$^{42.6}$~erg~s$^{-1}$, apparently in conflict with existing explosion models.}
    {}

\authorrunning{Sollerman, Yang, et al.}
\titlerunning{ZTF BTS SE SN luminosities}
\maketitle

\section{Introduction}
\label{sec:intro}

Core-collapse (CC) supernovae (SNe) are the final explosions of massive stars ($\gtrsim8~M_\odot$).
Hydrogen-poor SNe  represent CC in such stars that had lost most - or even all - of their envelopes prior to explosion. 
This includes Type IIb SNe (some H left), SNe Ib (no H, some He), and SNe Ic (neither H nor He). 
We will refer to these as stripped envelope (SE) SNe.

Even though SE SNe are 
less common than Type II SNe
\citep[e.g.,][]{li2011,graur2017}, 
there now exists a fair number of well observed objects. 
Presentations of such samples have highlighted how simple analytical models, such as 
that initiated by \cite{arnett1982}, provide reasonable matches with the observed light curves. 
Collecting sizeable samples, such exercises have revealed that the estimated ejecta masses are relatively low,
often seen as an argument for binary interaction playing a major role in the stripping of the progenitor stars
(\citealt{lyman2016}, \citealt{taddia2015sdss,taddia2018csp,taddia2019iptf}, \citealt{prentice2016,prentice2019}, \citealt{drout2011}, 
and \citealt{barbarino2020}). 

The other main result from these studies is that the amount of ejected radioactive nickel is typically larger than for normal Type II SNe. 
The mean value from the recent sample of Type Ic SNe from the iPTF survey \citep{barbarino2020} for example, concluded that 
M$_{^{56}\mathrm{Ni}} = 0.19 \pm 0.03~$\Msun.

A literature compilation by \cite{anderson2019}
calculated a median  
M$_{^{56}\mathrm{Ni}} = 0.032$~\msun\, for SNe II, and  
0.163 and 0.155~\msun\, for SNe Ib and Ic, respectively.
That study was repeated and augmented by
\citeauthor{2020A&A...641A.177M} (2020, see also \citealp{sharonkushnir2020MNRAS.496.4517S})
concluding that there 
exists a real, intrinsic difference in the amount of radioactive nickel between SNe II and SE SNe, even if the exact numbers are sensitive to the methodology.

Our paper takes two modeling studies as the starting point. Exploiting state-of-the-art neutrino-driven explosion models for  
 massive helium stars that have been evolved including mass loss, \cite{ertl2020}  
 note that for standard assumptions regarding the explosions and nucleosynthesis, their models predict light curves that are typically less luminous than many observed SNe Ib and Ic.  
 Their upper limit on the peak luminosity is 10$^{42.6}$ erg s$^{-1}$.  
 They remark that many 
 SNe Ibc appear to be 
 too luminous to be made by their neutrino-driven models, and propose that magnetars could be a promising alternative to power these supernovae, rather than, or in addition to, radioactivity.
 Alternatively, they suggest that observers could pay more attention to
 e.g., bolometric corrections, Malmquist bias or evidence for circumstellar interaction that could overestimate the reported peak luminosities.

 Following \cite{ertl2020}, \cite{woosley2020} augmented that study by adding detailed radiation transport. 
 Using the code SEDONA, they explored the same explosion models and could translate the limits on ejected nickel mass and bolometric luminosity 
 to maximum light in common filter pass bands.
 They have no models brighter than M$_r = - 17.8$ (or M$_g > -17.5$). 
 The bottom line in \cite{woosley2020} 
is that most SE SNe are best understood in 
\say{a traditional
scenario of binary mass exchange, neutrino-powered explosions without rotation, and radioactivity-illuminated
light curves}.  They thus seem less keen to lean on the magnetar solution, even though they acknowledge that
a sizeable fraction of the SE SNe might be out of reach (too bright) for their models.

 \cite{woosley2020} also occasionally discuss observational uncertainties, such as if some specific SNe might have had their host extinction over-estimated \citep[see also][]{Dessart2020A&A...642A.106D}, whether 
 some are really \say{normal} Type Ibc SNe, if the bolometric light curves (LCs) have been improperly assembled or if too simplistic modeling has been used to derive the amount of radioactive nickel.  
 They explicitly encourage observers to undertake new surveys and compare to their predicted pass-band LCs.
 Taking up that baton, our paper has a simple single goal in trying to address this question: Does a reasonable number of well-observed normal SNe Ibc reach peak luminosities in excess of
 M$_r = -17.8$ even if carefully assessing for e.g., distance and extinction? We explore which caveats such an investigation must consider.

 We make use of the sample of SE SNe (Type Ib and Ic, collectively labeled SNe Ibc) provided by the Zwicky Transient Facility \citep[ZTF,][]{GrahamM2019,Bellm2019}. 
 In particular, \cite{Fremling2020} introduced the ZTF Bright Transient Survey (BTS), which provides a large and purely magnitude-limited sample of extragalactic transients in the northern sky, suitable for detailed statistical and demographic analysis.  The early results of this survey were presented by \cite{Perley2020}, also introducing a web-based portal open to the public where specific sub samples can be constructed. We used this BTS sample 
 explorer\footnote{\href{https://sites.astro.caltech.edu/ztf/bts/explorer.php}{https://sites.astro.caltech.edu/ztf/bts/explorer.php}} 
 to collect all Type Ibc SNe within the BTS.
This is also an explicit purpose of this paper, to advocate the public BTS sample and to show how it can be used to address a specific scientific question. 

The paper is organized as follows. 
In Sect.~\ref{sec:obs} we present the observations and explain the sample selection based on our optical photometry and spectroscopy. 
Section~\ref{sec:discussion} presents a discussion of the different caveats in determining absolute magnitudes, including distances and extinction 
for this subsample.
Finally, Sect.~\ref{sec:conclusions} presents our conclusions 
and a short discussion where we put our results in context.

\section{Observations and sample}
\label{sec:obs}

\subsection{Survey and Selection of sample}
\label{sec:detection}

All photometric observations in this paper were conducted with 
the Palomar Schmidt 48-inch (P48) Samuel Oschin telescope as part of the ZTF survey, using the ZTF camera \citep{dekany2020}.
The light curves from the P48 come from the ZTF pipeline \citep{Masci2019}.
All magnitudes are reported in the AB system.

The BTS SNe are regularly reported to the Transient Name Server 
(TNS\footnote{\href{https://wis-tns.weizmann.ac.il}{https://wis-tns.weizmann.ac.il}}), and the LCs can be displayed using the above mentioned
BTS sample explorer, which we use to construct our sample.
We note again that the BTS is an untargeted
sample of SNe that is virtually spectroscopically complete down to a magnitude of 18.5
\citep{Perley2020}.

The aim of the paper is to explore to what extent there exist normal\footnote{There is no general definition of normal in this case, but in the rest of the paper we make sure the SNe are spectroscopically classified as Type Ib or Ic and not any peculiar sub-type, we reject SNe with unusual or broad LCs and with atypical colors.} 
Type Ibc SNe that exceed the maximum brightness predicted by the models mentioned in the introduction. Our main aim is therefore not to construct a complete and non-biased sample. Such a sample would of course be interesting to compare the average properties of SNe Ibc with the models, but would require greater care in terms of completeness and corrections for Malmquist bias \citep[see e.g.,][]{2021arXiv210900603O} and extinction. We take a simpler approach in this paper. Our aim is a reasonable number 
($\mathcal{O}(10)$) of normal SE SNe, large enough to not be biased by statistical outliers. More explicit investigations of the sample luminosity-function, light-curve parameters and extinction-correction properties are planned for future work.

Important for the selection is to have good enough data to construct the LCs, measure the peak luminosity, and ensure that the object is indeed a normal SN Ibc, both in terms of LC and spectra. 
In the first initial construction of the sample, we use the BTS explorer criteria provided in Table~\ref{tab:BTScriteria}.
The full BTS database included 
4496 objects classified as SNe, whereof 3038 were SNe that passed these cuts\footnote{Queried on 2021 June 28.}.
This included 
218 SE SNe.
The quality cuts in Table~\ref{tab:BTScriteria} ensure for example that our objects have data both before and after peak, and that the object was not detected too early in the survey when uncontaminated templates were not available.

From the initial list of 218 SE SNe we meticulously exclude candidates that do not fulfill the next sets of selection criteria. Since the BTS explorer includes $>200$ SNe Ibc, we can allow for rather strict cuts. 
These are based on data quality and are not supposed to bias the sample, more than in the requirement that the selected SNe are normal SNe Ibc. Note in particular that luminosity is not explicitly used in the sample cuts.
We further request that the classification Type is either Type Ib or a Type Ic. 
We thus remove all of the following types from the sample;
Types Ic-BL, Ibn,  Icn, IIb or Ib/c or Ib-pec, as well as anything labeled with a question mark.  
This excludes objects where other powering mechanism could be at play, such as shock cooling, circumstellar matter (CSM) interaction or a central engine. The "Ib/c" class on BTS represents objects for which a separation into either Type Ib or Type Ic could not be made based on the quality of the spectrum. For purity, we simply remove these from our sample as well. Finally, a few objects had different classifications on TNS as compared to our internal follow-up marshal. We removed these as well\footnote{This excluded the very luminous Type Ib/c SN\,2019jyn which clearly also challenges the conventional explosion models (Fraser et al., in preparation).}. 
This gave in the end 53 SNe Ib and 76 SNe Ic, or in total 129 Type Ibc SNe. 
The selection cuts are provided in Table~\ref{tab:Cuts}.

\begin{figure*}
\centering
    \includegraphics[width=0.8\textwidth]{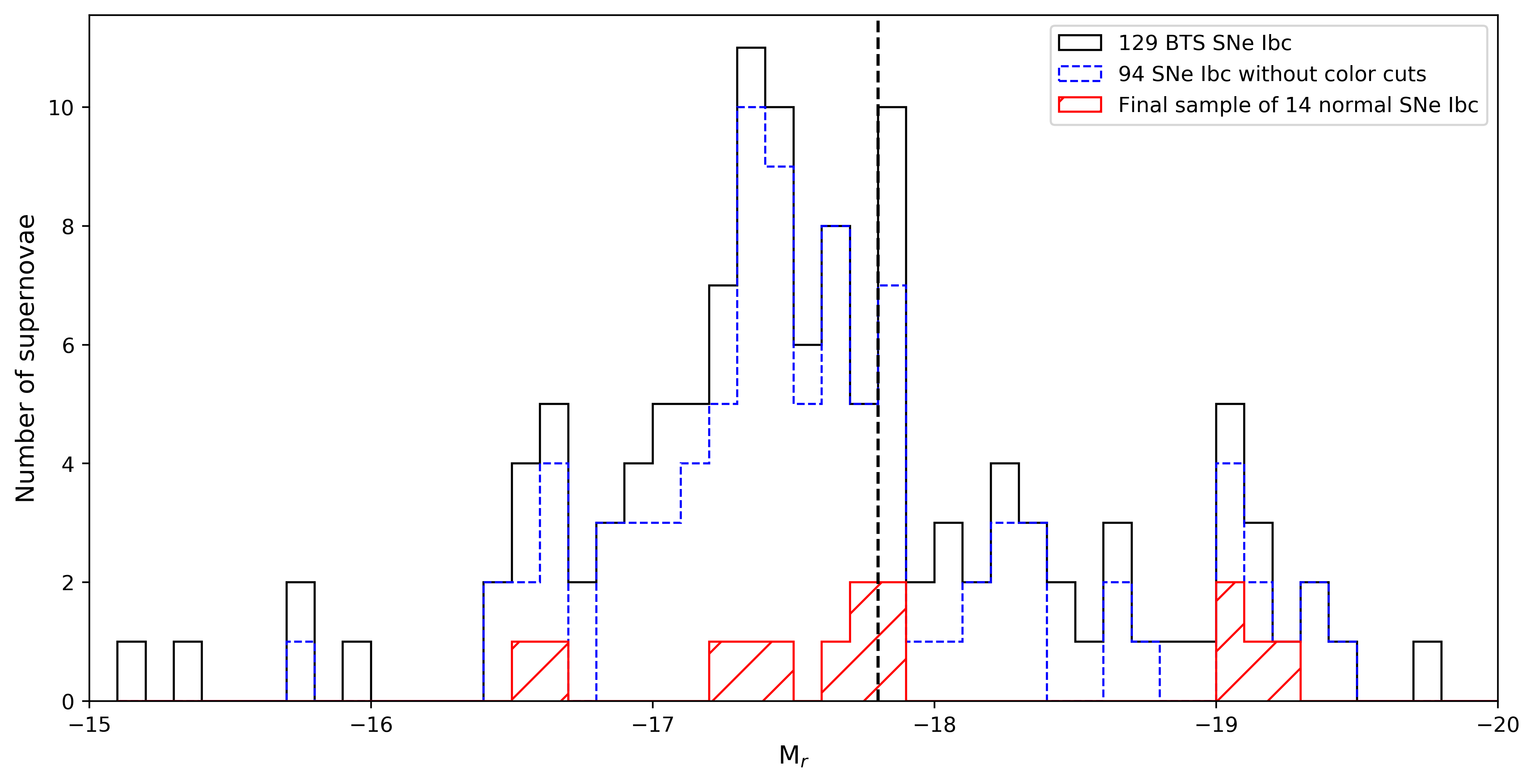}
    \caption{Luminosity function for Type Ibc SNe.
    The figure shows the number of objects per absolute magnitude bin (M$_r$) for different sample selections. The black distribution is for the 129 SNe Ibc initially selected from the BTS explorer and using the absolute magnitudes from that site. 
    This distribution has an average and standard deviation of M$_{g/r} = -17.61 \pm 0.72$ mag. 
    The blue dashed distribution is for the 
    94 
    SNe Ibc kept after additional quality cuts have been implemented. 
    These magnitudes are measured using GP on forced photometry data and yield M$_r = -17.64 \pm 0.54$ mag.   The red distribution of the final 14 
    normal SNe Ibc
    has an average and standard deviation of M$_r = -17.90 \pm 0.73$ mag.  
    The vertical black dashed line marks the upper limit of  M$_r = - 17.8$ from \cite{woosley2020}.
    }
    \label{fig:luminosityfunctionforfirstcuts}
\end{figure*}

This sample that fulfils our first set of BTS sample criteria is used to construct an initial luminosity function. 
The absolute peak luminosity function for these supernovae, with magnitudes as provided from the BTS, is presented in Fig.~\ref{fig:luminosityfunctionforfirstcuts} in black full lines. These BTS absolute magnitudes are computed using the observed peak, given the observed redshift and Milky Way extinction, and applies a basic k-correction.
This is already a significant result given the untargeted nature and the large size of the survey, and that the selection criteria used are mainly dependent on data quality and cadence. The sample and the luminosity function\footnote{This is perhaps more of a luminosity distribution, the observed magnitudes are not corrected for host extinction or Malmquist bias. Note that the BTS page provides either $g$- or $r$-band maximum magnitudes, whichever is brightest.}
is further refined throughout the rest of the paper.

\begin{figure*}
\centering
      \includegraphics[width=0.8\textwidth]{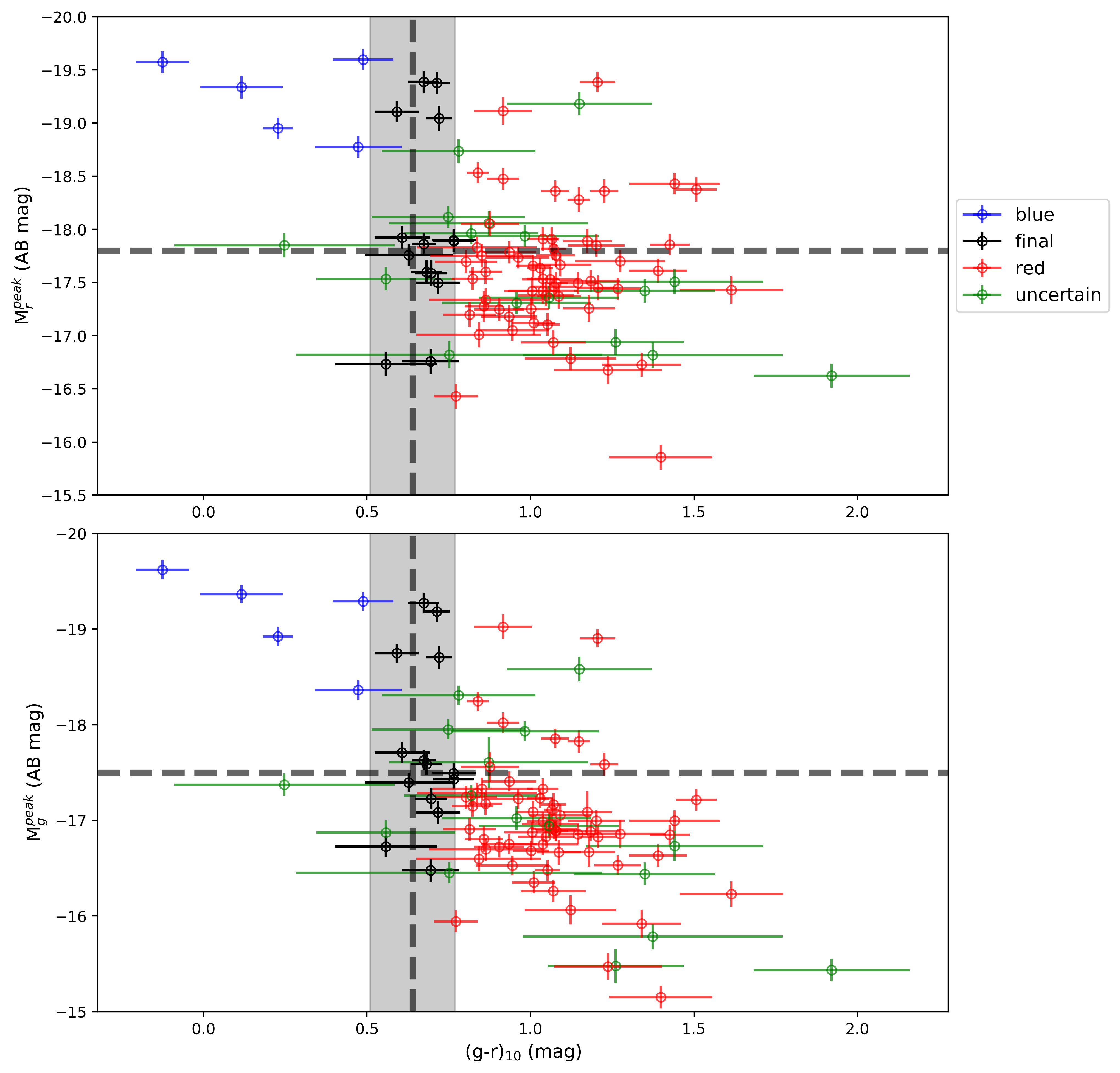}
    \caption{Colors and color cuts for the sample selections of Type Ibc SNe.
  The figure shows the absolute peak magnitudes (M$_r$ in the upper and M$_g$ in the lower panel) 
  for the 94 SNe selected versus their MW corrected colors in $g - r$ at $\sim10$ days after peak, when these transients have the most uniform color distribution \citep{taddia2015sdss,stritzingercolors}. The grey box includes those 14 SNe kept in the final sample as normal SNe Ibc where uncertainties in the extinction corrections are smaller. The color coding is explained in the main text. The horizontal dashed lines represent the maximum luminosities (M$_g = -17.5$, M$_r = -17.8$) according to \cite{woosley2020}. The data points also have uncertainties in magnitudes assigned, according to the error propagation in Sect.~\ref{sec:errors}.}
    \label{fig:colorsandcolorcuts}
\end{figure*}

\begin{figure*}
\centering
      \includegraphics[width=1.1\textwidth]{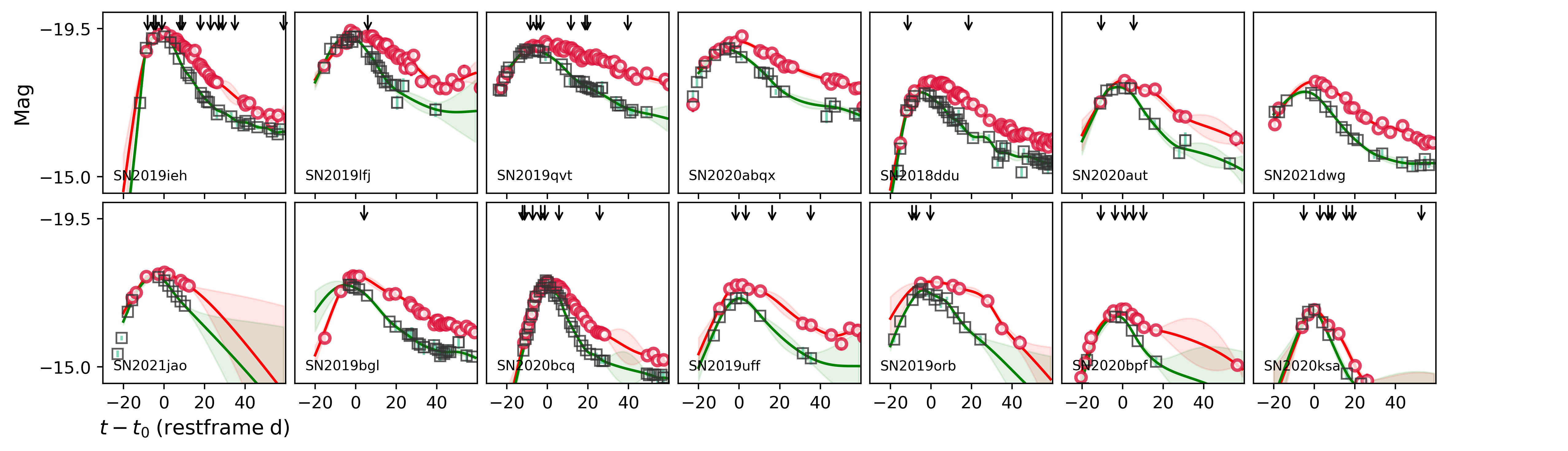}
    \caption{Light curves of the final sample of 14 Type Ibc SNe plotted in separate panels. We plot $g$- (green squares) and $r$-band (red circles) photometry in absolute AB magnitudes. These are corrected for distance and MW extinction. The x-axis gives rest frame days since $r$-band peak, where the redshifts and explosion dates are provided in Tables~\ref{table:SNproperties} and  \ref{table:SNLCproperties}. The dashed lines are the GP interpolations with error regions that were used to estimate peak explosion magnitudes and their uncertainties. Black arrows on top indicate the dates when we obtained spectra. 
    }
    \label{fig:lcs}
\end{figure*}

\subsubsection{Photometry cuts}

We proceed with those SNe that have good quality light curves. At this stage, we performed forced photometry \citep{Masci2019,yuhan2019} for the remaining SE SN subsample. 
For those resulting LCs, we furthermore require the following data quality cuts:
\begin{itemize}
    \item At least 6 epochs of photometry in either $g$ or $r$ band.  
    \item At least 3 epochs of $g-r$ (sampled within $\pm3$ days).
     \item Photometry available both before and after peak within $\pm 3$ days of estimated time of peak brightness. 
     \item Photometry accurate enough so that we can determine the peak luminosity to better than 10\% (0.1 mag). 
\end{itemize} 

These steps were accomplished using a 
Gaussian Processing (GP) algorithm\footnote{\href{https://george.readthedocs.io}{https://george.readthedocs.io}}
to interpolate the photometric data. 
The number of SNe that remains after each sample cut is presented in Table~\ref{tab:Cuts}.
We note again that selecting on cadence and data quality should not bias the sample in preferring some specific 
classes of SE SNe before others, or deselecting particular environments. There is, however, a Malmquist-like selection in that intrinsically very faint or fast transients will on average have fewer good-quality datapoints. For the purpose of this study of the bright end of the luminosity function, where we want to find out if there exist luminous SNe, this is not a problem -- but we note that there may exist a population of less luminous, nickel-poor SE SNe that are underrepresented or missing from this compilation. 
Fremling et al. (in prep.) are exploring ways to find such transients by their early shock-breakout cooling emission.
The rationale for requiring two bands at this stage is that we also want to be able to construct bolometric LCs and to assess the host extinction, see Sect.~\ref{sec:bolcor} and \ref{sec:extinction}.
Only 10
objects were removed in this step, mostly since we had already performed cuts on the data in the first selection (Table~\ref{tab:BTScriteria}).

\subsubsection{Distance cuts}\label{sec:distancecuts}

Distances are often a major uncertainty in estimates of absolute luminosities and thus nickel masses. 
This is paradoxically often true for the most nearby, and therefore best observed, SNe in the literature - 
simply because in the local universe the peculiar motions of nearby galaxies make the relative distance uncertainties larger. To avoid SNe with large uncertainties from their distance estimates we 
require that the SNe 
are distant enough to be in the Hubble flow (we use $z > 0.015$). None of the nearby hosts had a distance estimate from e.g., Cepheids. This excludes seven rather well observed SNe\footnote{ZTF20aaelulu, ZTF20acpjqxp, ZTF21abcgaln, ZTF20aavzffg, ZTF21aaqhhfu, ZTF21aaxtctv, ZTF21aaaadmo.} from this study.

Redshifts were converted to distances using a
flat cosmology with H$_0=70$~km~s$^{-1}$~Mpc$^{-1}$ and $\Omega_{\rm{m}} = 0.3$.
The rationale for this cut and the
remaining uncertainties from these distance estimates to the absolute magnitudes 
are further discussed in Sect.~\ref{sec:distances}.

\subsubsection{Milky Way reddening}\label{sec:mwreddening}

In our analysis we correct all photometry for Galactic extinction, using the Milky Way (MW) color excess 
$E(B-V)_{\mathrm{MW}}$ toward the position of the SNe, as provided in Table~\ref{table:SNproperties}.
These are all obtained from \cite{2011ApJ...737..103S}. All reddening corrections are applied using the \cite{1989ApJ...345..245C} extinction law with $R_V=3.1$. Supernovae experiencing significant amount of Galactic extinction (A$_{\rm{V}} > 1.0$ mag) were already deselected in the BTS explorer search (Table~\ref{tab:BTScriteria}). For this exercise, we furthermore remove SNe for which the MW extinction correction A$_{\rm{V}} > 0.5$ mag, see Table~\ref{tab:Cuts}. The argument is simply that larger corrections imply larger uncertainties. The corrections for dust in the host galaxies is discussed below (Sect.~\ref{sec:extinction}). This removed six objects (Table~\ref{tab:Cuts}).

\subsubsection{Light curve properties}
\label{sec:lightcurveproperties}

To make sure we select only normal SE SNe, since these are what we want to compare against, we made LC fits using a functional form used for SNe also in \citet[][see their fig.~8]{taddia2015sdss}.  
This was done using {\tt scipy.optimize.curve.fit} 
and we require the fit to have $\chi^{2} < 2$ per degree of freedom. 
This step is made to avoid SNe with LC bumps, signs of CSM interaction, or just too poor photometry. 
This removed only a few 
SNe\footnote{Including the double peaked SN 2019cad \citep{claudia}, and the unusual SN 2018ijp \citep{leonardo2021}.}.

In this exercise we also use the LC fit with the analytical function to characterise the rise and decline parameters \cite[again following the study of][]{taddia2015sdss}. In order to estimate the actual rise time with respect to an estimated time of explosion (first light), we followed the methodology employed by \cite{MillerZTFIa} using both the pre-explosion upper limits and the rising part of the LC. 
Comparing the $\tau_{\rm{fall}}$ vs $\tau_{\rm{rise}}$ distributions with those of the SDSS sample \citep{taddia2015sdss}, and in particular investigating the rise-time distribution, we decided to remove objects with $\tau_{\rm{rise}} > 8$ days. This effectively also removed all objects with t$_{\rm{rise}} > 35$ days\footnote{Whereas $\tau_{\rm{rise}}$ measures how fast the LC rises pre-peak according to the formalism of \cite{bazin},  t$_{\rm{rise}}$ measures the actual time from estimated explosion time to peak luminosity.}. 
Again, the selection is made to focus this study on the normal population of SNe Ibc. Slow-rising SE SNe are by themselves also of large interest, in particular for understanding the population of single massive stars as progenitors, but for the scope of this investigation such objects are de-selected.

Out of the initial 129 SE SNe, 94 
remained after the above mentioned cuts. The absolute peak luminosity function for these supernovae is also presented in Fig.~\ref{fig:luminosityfunctionforfirstcuts} (dashed blue lines). 
This is a significant contribution to the knowledge of the Type Ibc luminosity function, and the sample compares well with for example the recently published large iPTF sample of 44 SNe Ic by \cite{barbarino2020}, but with a higher degree of control on the selection functions.
The results will be discussed further in the next sections, but for now we proceed to a final culling of our sample.

\subsubsection{Host galaxy extinction}

This final cut is made to remove objects with different colors than the main population of SNe Ibc. The main rationale here being that we want to avoid large corrections for host-galaxy extinction. This is 
probably the largest uncertainty that could be ingested from the observational side, over-correcting for extinction would make the SNe too luminous, which could be a reason for the apparent discrepancy between model predictions and observations.

A very red color for the MW-extinction corrected SN LC probably indicates significant host-galaxy extinction. There are a number of ways to compensate for this, as discussed in Sect.~\ref{sec:extinction}, but all of the methods come with a (fairly large) degree of uncertainty.

To exclude cases where extinction corrections would come with a large uncertainty, we simply deselect objects that are too red ($g-r > [0.64+0.13]$ mag) at 10 days past peak, and also cut out objects that are significantly bluer ($g-r < [0.64-0.13]$ mag) than the rest of the sample at this phase. These numbers and their uncertainties come from the investigation of \cite{taddia2015sdss} using Type Ibc SNe from SDSS.
We furthermore reject objects where the color information is simply not accurate enough to reliably perform these cuts, i.e. we reject any object for which we can not estimate ($g-r$) at 10 days past peak with an accuracy better than 0.2 mag. This is clearly one of the most severe cuts in the sample selection, removing 59+5+16 objects, where only 14 remain (Table~\ref{tab:Cuts}). The rationale for these cuts and the remaining uncertainties are further discussed in Sect.~\ref{sec:extinction}. The selection is illustrated in Fig.~\ref{fig:colorsandcolorcuts} where
the grey area shows the typical colors of SE SNe at 10 days past peak, $g-r = 0.64\pm0.13$ mag \citep{taddia2015sdss}. The objects that survive this final cut are marked with black symbols. The red symbols constitute the majority of the objects, which have redder colors. The notion that they are also dimmed by extinction is supported by the fact that they are typically less luminous than the bluer SNe; there is a clear trend visible in this figure. Instead of attempting to correct for this dimming, for the final cut in this paper we conservatively simply remove all of these objects. We stress that this is very cautious with respect to the purpose of this study, the red objects would only become more luminous with host extinction corrections (Sect.~\ref{sec:redsample}).
The green symbols in Fig.~\ref{fig:colorsandcolorcuts} show the objects that were removed because the GP photometry at +10 days had too large uncertainties on the color. 
Finally, we note a sub-population of bright and blue objects, marked with blue symbols in the upper left corner of the figure. Including these objects would also make our average SN Ibc magnitude brighter, and the required mass of radioactive material larger. Conservatively, we remove them on the basis that they do not have normal colors according to \cite{taddia2015sdss} and \cite{stritzingercolors}.
The final selection leaves us with only 14 SNe. The properties of these SNe, with regards to the selection criteria detailed above, are provided in Table~\ref{table:SNproperties}.

{ Most of the spectroscopy for the BTS is conducted with  the robotic 
Palomar 60-inch telescope (P60; \citealp{2006PASP..118.1396C}) equipped with the Spectral Energy Distribution Machine (SEDM; \citealp{2018PASP..130c5003B}). Further spectra were often obtained with other larger telescopes such as the Nordic Optical Telescope (NOT) using the Alhambra Faint Object Spectrograph (ALFOSC). 
The BTS provide all classification spectra to the publicly available TNS. SEDM spectra were reduced using the pipeline described by \citet{rigault}. 
} We have checked the spectra for these objects, and confirm that they are all best fit with the reported subclasses of SNe.

\section{Discussion}\label{sec:discussion}

The properties of the final sample of SNe Ibc 
are listed in Table~\ref{table:SNproperties}.
The ZTF obtains regular photometry in $g$, $r$ (and $i$) bands, 
and for these 14 SNe we have on average 13/14 $r/g$-band points between $-20$ and 20 days past peak, i.e a typical cadence of 3 days, although some are better sampled than 2 days.
The SNe in this sample thus have 
relatively well-constrained explosion times, 
rise times and decline times, and we measure these parameters and list them in 
Table~\ref{table:SNLCproperties}.

In Fig.~\ref{fig:lcs} we show their LCs in absolute magnitudes. The magnitudes are in the AB 
system and have been corrected for distance modulus and MW extinction.
They are plotted versus rest frame days past estimated explosion epoch.  
This final absolute magnitude distribution is included in Fig.~\ref{fig:luminosityfunctionforfirstcuts}. 

Next we briefly discuss some of the selection cuts and the corrections and their uncertainties, given the main aim of this investigation. We make an effort to quantify the uncertainties involved in the different steps, to be able to propagate these to the final luminosity function.

\subsection{Distance estimates}
\label{sec:distances}

An important uncertainty in estimating absolute luminosities (and nickel masses) for SNe is the uncertainties in the distance estimates.  Such uncertainties are often under-appreciated in the SN literature. In particular, many studies focus on nearby objects where good data quality is easier to acquire, but where the relative uncertainties due to peculiar motions of the host galaxies can be considerable. 

As an example, we mention SN\,2020oi (ZTF20aaelulu), 
a nearby Type Ic SN that was part of our initial BTS sample of SNe Ibc. 
For SN~2020oi in the host galaxy M100, \cite{horesh2020} adopted a distance of 14 Mpc, 
corresponding to a distance modulus of $30.72 \pm 0.06$ mag.
For an Arnett type of model, the nickel mass basically scales linearly with peak luminosity
and a distance modulus uncertainty of 0.06 mag translates to a relative uncertainty on the ejected nickel mass of $5.5\%$.

However, the NASA Extragalactic Database (NED\footnote{http://ned.ipac.caltech.edu/}) 
includes multiple different distance estimates for this nearby host (and many others). Following the apporach of \cite{steer2020},   
an uncertainty from median combining many of those estimates would be 
$16.4\pm2.35$ Mpc, which would correspond to an uncertainty in the nickel mass of $29\%$.
To illustrate this, we note that the second published paper on 
SN\,2020oi use a distance of 16.22 Mpc \citep{2021ApJ...908..232R}. They quote a nickel mass with an uncertainty of 15\%, but we note that the difference in distance as adopted by these two studies of the same SN amounts to a 33\% difference in flux. Somewhat ironically, the studies reach similar conclusions since they also adopt different amounts of host extinction, which in this case happens to work in the direction of decreasing the differences. Note that SN\,2020oi would also have been deselected from our sample due to the large host extinction, which makes it difficult to accurately determine the intrinsic luminosity.

ZTF is an untargeted survey. Therefore, in contrast to most previous samples of SE SNe, we are not biased towards the nearby and massive galaxies. The redshift distribution of our (94 object) sample
has  a mean value and rms of $0.036\pm0.003$, which means that peculiar velocities for the host galaxies are of less importance. Estimating a typical peculiar velocity of 300~km~s$^{-1}$ \citep{Davis2011}
means that for our cut-off value 
$z=0.015$ we have an uncertainty on $cz$ of $<7\%$
whereas for the mean redshift of the 
sample ($\overline{z}=0.036$ within the errors for the three samples)
this gives a typical flux error of 3\%.
For our distance estimate uncertainties for the individual SNe in the final sample
we use an individual uncertainty from peculiar velocities of
150~km~s$^{-1}$ 
and for the cosmology we include a systematic uncertainty of $\pm3$~km~s$^{-1}$~Mpc$^{-1}$ on the Hubble constant (Sect.~\ref{sec:errors}).

\subsection{Host extinction}
\label{sec:extinction}

Correcting for host extinction is probably the most difficult part in determining the luminosity function for any type of SN. \cite{barbarino2020} used two different approaches for their recent SN Ibc sample, both from narrow absorption lines of \ion{Na}{I~D} in the spectra, and by using the SN colors to correct for reddening. There are pros and cons with both of these, and they are certainly both affected by uncertainties. Overall, on a sample level, the main results of \cite{barbarino2020} were not much affected by the choice of method, but for the individual SN the actual correction can vary substantially. It is widely accepted that there is some relation between deep host-galaxy sodium absorption lines and the amount of extinction, but the scatter is large and the implementations differ 
(\citealp[e.g.,][]{Turatto2003}, \citealp{Poznanski2012}, \citealp{Blondin}, and \citealp{Phillips2013}). 
For the ZTF SNe we have often rather low-quality spectra, and we will not adopt these methods.

The other methodology is to make use of the fact that SNe Ibc often have similar colors at some phase after peak. 
This was first noted by \cite{drout2011}, implemented by \cite{taddia2015sdss}
and further developed by \cite{stritzingercolors}.
The basic assumption here is the uniformity of these events, and interpreting redder events as being affected by host galaxy extinction. 
The investigations of \cite{stritzingercolors} and \cite{taddia2015sdss} define a range of colors for normal, un-reddened SNe Ibc, and we have adopted the cuts from the latter study\footnote{The typical colors from \citet[][their table 2]{stritzingercolors} are slightly redder, $g - r \sim0.8$ mag, and we stick to the bluer estimate from \cite{taddia2015sdss}.} on the uncertainties and actual colors at 10 days past peak (Fig.~\ref{fig:colorsandcolorcuts}, Table~\ref{tab:Cuts}).

However, in this paper we remain cautious on the actual and quantitative host-reddening correction. Our conservative approach is therefore to not apply any correction for host galaxy reddening, and simply remove the objects for which such a correction would have been needed. This culls a large fraction of our sample, but also alleviates the main problem. Figure~\ref{fig:colorsandcolorcuts} illustrates the situation, where absolute magnitudes in the $r$ band (M$_{{r}}$, top) and $g$ band (M$_g$, bottom) are plotted versus MW corrected colors at 10 days past peak. The black vertical line shows $g-r = 0.64$ mag which is the normal unreddened color for SNe Ibc, and the grey region shows the $1\sigma$ deviation on this number ($\pm0.13$ mag) from \cite{taddia2015sdss}.
The red symbols show the large fraction of SNe that have redder colors and are therefore suspected to be affected by host galaxy reddening. These are excluded from the final sample. On the left-hand side of the gray region there are also a number of SNe (5) that have bluer colors than the typical SN Ibc. These are marked with blue symbols and are also de-selected (Sect.~\ref{sec:extinction}, Table~\ref{tab:Cuts}). 

We note that there is indeed a correlation between absolute magnitude in these two bands and color at 10 days. The slope of the correlation is also larger in the $g$ band, as expected if this is primarily due to extinction by dust. 

\subsection{Luminosities and bolometric corrections}
\label{sec:bolcor}

As a final exercise, we attempt to construct bolometric LCs for our final sample and use analytic expressions to estimate the amount of radioactive $^{56}$Ni needed to power the peaks of these LCs.
We follow the procedure outlined by \cite{LymanBC} in order to construct the bolometric LCs from the 
$g$- and $r$-band LCs. This is a well established procedure for normal Type Ib and Type Ic SNe, and we have secured that our final objects constitute such a sample. 
We thereafter estimate the nickel-mass following a simple Arnett model \citep{arnett1982,leonardo2021}.
This provides final bolometric luminosities with
corresponding nickel masses of 
M$_{\rm{Ni}} = 0.25 \pm 0.05$~\msun\ 
for the sample of 14 SNe Ibc. This compares well with the values from the investigation of \cite{barbarino2020}, with 
M$_{\rm{Ni}} = 0.19 \pm 0.03$~\msun~for 41 SNe Ic, which used a similar approach. 
The larger sample of 94 SNe Ibc has a mean value of  M$_{\rm{Ni}} = 0.16$ \msun, but we note that no host-extinction corrections were applied to that sample. There is an ongoing discussion on to what extent the simple models used here infer a realistic nickel mass, and other alternatives have been suggested 
(\citealp{Dessart2016MNRAS.458.1618D}, \citealp{KKnimass}, and \citealp{Afsariardchi2021ApJ...918...89A}). 

This is the reason why we mainly stick to the pass-band magnitudes in this observational paper, to directly compare with the predictions from the radiation transport of \cite{woosley2020}.
Work to estimate nickel masses for this sample using mutliple methods is ongoing.

\subsection{Error propagation}
\label{sec:errors}

Apart from presenting mean values and rms uncertainties on the absolute magnitudes for the sample populations, we have also propagated the uncertainties for the individual objects 
through the different steps as outlined above. For each individual supernova we include the
photometric uncertainty on the peak magnitude as estimated from our GP analysis, a 15\% uncertainty in the correction for MW extinction, a 150~km~s$^{-1}$ uncertainty included in the peculiar velocity correction and a systematic $\pm~3$~km~s$^{-1}$~Mpc$^{-1}$ error on the adopted Hubble constant. These uncertainties are then provided as error bars on the y-axis for the black symbols in Fig.~\ref{fig:colorsandcolorcuts}. These magnitude errors are mostly $<15$\%.

\subsection{Final sample}
\label{sec:finalsample}

The sample criteria so far have been strict and objective, without dwelling on any individual SN. The three sample distributions in Fig.~\ref{fig:luminosityfunctionforfirstcuts} actually all have the same mean values within the errors, but the final sample is somewhat limited by statistics and four objects at M$_{r} \sim -19$ are significantly more luminous than the model limit whereas there are no objects left in the bin between $-18$ and $-19$.  Here we briefly review first the four final objects that are substantially brighter than the model limits. Thereafter we also look individually on four objects with 
M$_{r} \sim -18.5$ which were de-selected due to their red colors, and discuss if these are also robustly brighter than the investigated limit magnitude.

\begin{figure*}
\centering
     \includegraphics[width=0.8\textwidth]{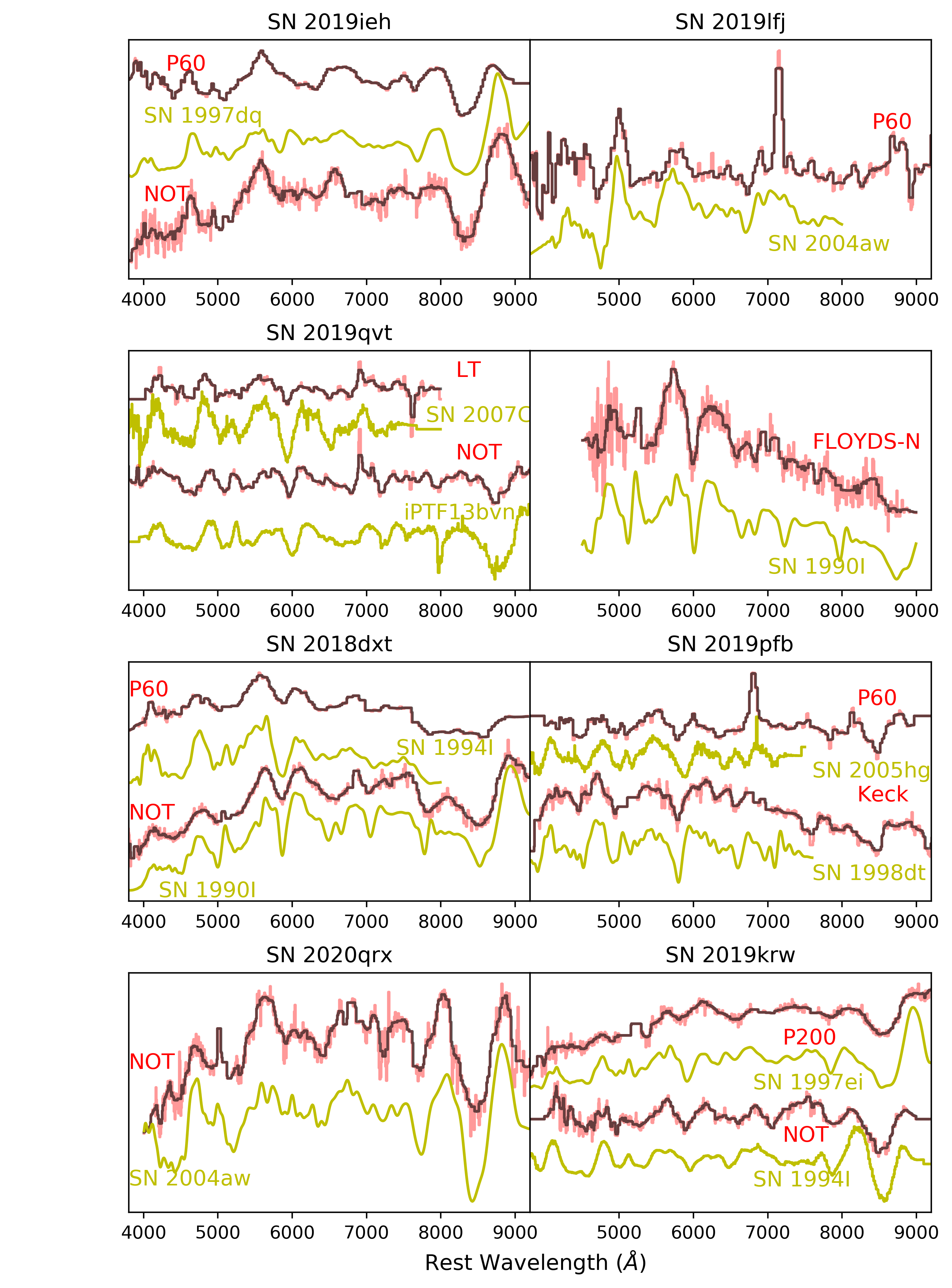}
    \caption{Spectra for the bright and the red SNe discussed in Sect.~\ref{sec:bright} and Sect.~\ref{sec:redsample}. We display the classification spectrum from TNS together with a named template spectrum to illustrate the quality of our spectra. For most objects we also have additional spectra (see Fig.~\ref{fig:lcs}), and we provide some of these here as well. All spectra are made available on Wiserep.
    Our spectra are shown in red, where the black line is a smoothed version. The name of the used telescope is also provided.
    The classification spectrum for SN 2019pfp is from \cite{2019TNSCR1749....1T} and the one from
    SN 2020abqx from \cite{2021TNSCR..50....1B}.
    }
    \label{fig:spectra}
\end{figure*}

\subsubsection{The bright ones}\label{sec:bright}

Four objects in the final sample have absolute magnitudes brighter than $-19$. This is substantially more luminous than the predictions from the explosion models, and also on the bright end of the entire luminosity distribution. Although there are also several SNe robustly around $-17.8$ which are challenging the predictions, we first individually look at the top four. Treating samples on an overall statistical level is certainly more objective, whereas scrutinizing individual objects can illuminate some of the sample caveats. 
{ For these objects we also display the classification spectra in Fig.~\ref{fig:spectra}. For most objects we have multiple spectra to support the classification (as indicated on top in Fig.~\ref{fig:lcs}).  }

\begin{itemize}
\item
SN 2019ieh / ZTF19abauylg:

This SN has a well monitored LC and 
the redshift is secure from host galaxy emission lines in the SN spectrum.
{ The classification spectrum is from P60 and is shown in Fig.~\ref{fig:spectra}. A better spectrum was obtained with the NOT, and is best fit by a Type Ic template using SNID \citep{Blondin}.} The same is true for an earlier Lick spectrum, although templates with SNe Ic-BL are also viable fits at that phase.

\item
SN 2019lfj / ZTF19abfiqjg:

The redshift is secure from the host galaxy spectrum (SDSS). The LC is well sampled in the $r$ band and peaks above $-19$; it is poorly matched with SN Ia LC using SALT2 \citep{salt2}. The classification is based on a single host-contaminated P60 spectrum. 
{ The best SNID fit template is the Type Ic SN 2004aw.}

\item

SN 2019qvt / ZTF19abztknu:

The redshift is also in this case from host lines in late SN spectra. 
The LC is well sampled and the
Type Ib classification is secure from He lines in later spectra. 
{ In Fig.~\ref{fig:spectra} we display both the classification spectrum obtained with the Liverpool Telescope, as well as a later NOT spectrum, together with matched Type Ib template spectra.}

\item
SN 2020abqx / ZTF20acvebcu:

For this SN Ib, the classification spectrum from 
\cite{2021TNSCR..50....1B} includes also galaxy lines, securing the redshift (host $z$ also known from SDSS). Also SNID finds good matches with a SN Ib at this redshift {(Fig.~\ref{fig:spectra})}. The classifiers note that the \ion{He}{I}~$\lambda5876$ is particularly strong. We do not have a detailed spectroscopic sequence to establish the classification further. The LC is not well fit with a Type Ia SN LC with a secondary $r$-band bump. 

\end{itemize}

These are thus clearly luminous supernovae with secure peak photometry and redshifts. Some of the objects have classifications based on low-resolution and mediocre signal-to-noise spectra from robotic telescopes, where the potential confusion could be with peculiar SNe Ia or Type Ic-BL SNe.

\subsubsection{The red ones}\label{sec:redsample}

We also discuss four objects with brightness significantly above the theoretical limit, but which were excluded from the final sample because they were slightly too red in Fig.~\ref{fig:colorsandcolorcuts}\footnote{ZTF18abfzhct, ZTF19abvdgqo, ZTF20abqdkne, ZTF19abdoior.}.
Looking also at these individual objects, we conclude that they are all positioned in ({projected on}) large star-forming galaxies which is consistent with them suffering from some extinction. For three out of four there were no previous host galaxy redshifts, but our later spectra secure these from host galaxy lines. Again the light curves are well  sampled and there are no doubts on redshifts or photometry. In all cases early robotic spectra are complemented with later spectroscopy from larger facilities, and also here we have no reason to reclassify any of the objects. {Some of these spectra are also displayed in Fig.~\ref{fig:spectra}. We show the classification spectra from TNS and the closest template according to SNID. In some of these cases there exist also earlier P60 spectra, but these were not of high enough quality to warrant a secure classification.
}
These are thus SE SNe brighter than the investigated limits, and any corrections for host galaxy extinction would only make them even brighter.

The conclusion from investigating these individual SNe in the sub-sections above is that for some of the objects the exact sub-classifications might be questioned, but that overall we often have multiple spectra and supporting observations also from larger facilities. Redshifts derived from the supernova features may come with larger uncertainties, but for the objects investigated here all redshifts were well established from host emission lines. There are thus clearly normal SE SNe that reach above the brightness limits investigated in this study. 
It is also clear that many of the red objects are normal SNe Ibc that are already brighter than the limit. For the sample of 94 objects, there were 29 such SNe (31\%). Correcting for host extinction only makes that large sample brighter.

\section{Summary and conclusions}\label{sec:conclusions}

In this paper we have presented the SE SNe from the BTS sample. Starting with 129 selected Type Ib and Type Ic SNe from the BTS, we present a first luminosity distribution for these objects. This is shown in Fig.~\ref{fig:luminosityfunctionforfirstcuts}. 
The mean absolute magnitude and the rms for this distribution is 
M$_{\rm{g/r}} = -17.61 \pm 0.72$, and 36\% of the SNe appear brighter than the limit of 
$-17.8$ that \cite{woosley2020} suggested as the upper limit on the brightness from their radiation transport calculations based on state-of-the-art explosion models. This already supports previous studies reporting large luminosities and nickel masses for Type Ibc SNe.

A main driver in this paper has been to use the well characterised BTS sample together with strict selection cuts to weed out the normal SNe Ibc. One of the largest cuts in the selection of the final sample was on the colors of the SNe. This was discussed in Sect.~\ref{sec:extinction} and illustrated in Fig.~\ref{fig:colorsandcolorcuts}. 
Correcting for extinction would make the red objects to the right even more luminous, further amplifying the discrepancy between the model predictions and the observed luminosity function. Several of these bright and red objects are clearly SE SNe more luminous than the theoretical cut (Sect.~\ref{sec:redsample}).
We also note the objects marked in blue that we have also de-selected from the sample. The rationale for omitting these objects was not that they are affected by dust, but merely that they are outside the region of normal SN Ibc colors. It is noteworthy that they are all more luminous than M$_{\rm{r}} = -17.8$. Including some of these objects would clearly push the luminosity function to even brighter magnitudes. Similarly, declaring some of them as normal, un-distinguished SNe would effectively push the black vertical line to the left, and also make the final sample more luminous.

We have used the ZTF BTS sample and a series of selection criteria to investigate if normal SE SNe can be more luminous than M$_{{r}} = -17.8$. They can!
This puts the ball back on the theoretical model court, implying either modifications to the fundamental core-collapse explosion models, alternative powering mechanisms (such as magnetars), more sophisticated radiative transport schemes to translate bolometric luminosities to pass-band limits, or probably a combination of these.

\begin{acknowledgements}
We thank Schuyler van Dyk for comments.
{ Thanks to the referee for encouraging comments.}
Based on observations obtained with the Samuel Oschin Telescope 48-inch and the 60-inch Telescope at the Palomar Observatory as part of the Zwicky Transient Facility project.  ZTF is supported by the National Science Foundation under Grant No. AST-2034437 and a collaboration including Caltech, IPAC, the Weizmann Institute for Science, the Oskar Klein Center at Stockholm University, the University of Maryland, Deutsches Elektronen-Synchrotron and Humboldt University, the TANGO Consortium of Taiwan, the University of Wisconsin at Milwaukee, Trinity College Dublin, Lawrence Livermore National Laboratories, and IN2P3, France. Operations are conducted by COO, IPAC, and UW. 
SED Machine is based upon work supported by the National Science Foundation under Grant No. 1106171. 
The ZTF forced-photometry service was funded under the Heising-Simons Foundation grant 12540303 (PI: Graham). 
 This work was supported by the GROWTH project \citep{2019PASP..131c8003K}
funded by the National Science Foundation under PIRE Grant No 1545949. 
The Oskar Klein Centre was funded by the Swedish Research Council.
Gravitational Radiation and Electromagnetic Astrophysical Transients (GREAT) is funded by the Swedish
Research council (VR) under Dnr 2016-06012. 
Partially based on observations made with the Nordic Optical Telescope, owned in collaboration by the University of Turku and Aarhus University, and operated jointly by Aarhus University, the University of Turku and the University of Oslo, representing Denmark, Finland and Norway, the University of Iceland and Stockholm University at the Observatorio del Roque de los Muchachos, La Palma, Spain, of the Instituto de Astrofisica de Canarias.
Some of the data presented here were obtained with ALFOSC. 
MMK acknowledges generous support from the David and Lucille Packard Foundation.
\end{acknowledgements}

\onecolumn

\begin{deluxetable}{l c}
\tablewidth{0pt}
\tabletypesize{\scriptsize}
\tablecaption{BTS sample explorer criteria\label{tab:BTScriteria}}
\tablehead{
\colhead{Criteria} &
\colhead{Fulfilled}
}
\startdata
Quality Cuts: & \\
Require pre/post peak coverage & Yes \\
Require good visibility      &  No \\
Require passes 2020B filter  & Yes \\
Require uncontaminated reference & Yes \\
Require peak after May 2018  & Yes \\
Require low extinction       & Yes \\
Purity cuts: & \\
Require SN-like light curve  &  Yes \\
Require galaxy crossmatch   &  No 
\enddata
\end{deluxetable}

\begin{deluxetable}{clc}
\tablewidth{0pt}
\tabletypesize{\scriptsize}
\tablecaption{Sample cut criteria
\label{tab:Cuts}}
\tablehead{
\colhead{Step} &
\colhead{Criteria} &
\colhead{Number of SNe}
}
\startdata
1 & Full BTS SN sample (from June 28 2021) & 4496 \\
2 & Full sample after criteria in Table~\ref{tab:BTScriteria} & 3038 \\
3 & SE SNe & 218 \\
4 & Type Ib (53) or Type Ic (76) & 129 \\

\hline
\multicolumn{3}{c}{-- Extract forced-PSF photometry light curves -- }\\
\multicolumn{3}{c}{SNR = 5 sigma}\\
\hline
5 & Data quality cuts & 119 \\
6 & Distance cuts, $z > 0.015$ & 112 \\
7 & MW A$_{\rm{V}} < 0.5$ mag &  106 \\
8 & LC template comparison & 94 \\
9 & Color cuts: & \\
& ($g-r$)$_{10} < 0.77$ mag & 89 \\
& ($g-r$)$_{10} > 0.51$ mag & 30 \\
& $\Delta$($g-r$)$_{10} < 0.2$ mag  & 14 \\
\enddata
\end{deluxetable}

\begin{deluxetable}{lccccccccccc}
\tablewidth{0pt}
\tabletypesize{\scriptsize}
\tablecaption{Final sample of supernovae and their host galaxies \label{table:SNproperties}}
\tablehead{
\colhead{ZTFID} &
\colhead{IAUID} &
\colhead{Type} &
\colhead{R.A. (SN)} &
\colhead{Decl. (SN)} &
\colhead{$z$} &
\colhead{$A_{V,\,\rm MW}$} &
\colhead{R.A. (host)} &
\colhead{Decl. (host)} &
\colhead{$M_{g,\,\rm host}$} & \\
\colhead{} &
\colhead{} &
\colhead{} &
\colhead{(hh:mm:ss)} &
\colhead{(dd:mm:ss)} &
\colhead{} &
\colhead{(mag)} &
\colhead{(hh:mm:ss)} &
\colhead{(dd:mm:ss)} &
\colhead{(mag)}
}
\startdata
ZTF19abauylg     & SN\,2019ieh  & SN Ic & 16:42:10.83 & +06:59:02.4 & 0.032 & 0.28 & 16:42:10.80 & +06:59:02.1 & $-17.33^{+0.13}_{-0.06}$\\
ZTF19abfiqjg     & SN\,2019lfj  & SN Ic & 01:57:48.75 & +13:10:34.2 & 0.089 & 0.15 & 01:57:48.76 & +13:10:34.4 & $-20.68^{+0.16}_{-0.04}$\\
ZTF19abztknu     & SN\,2019qvt  & SN Ib & 03:09:01.53 & +24:02:38.1 & 0.053 & 0.48 & 03:09:01.52 & +24:02:39.2 & $-18.56^{+0.13}_{-0.06}$\\
ZTF20acvebcu     & SN\,2020abqx & SN Ib & 11:52:24.66 & +67:32:51.7 & 0.063 & 0.03 & 11:52:24.94 & +67:32:51.8 & $-19.62^{+0.11}_{-0.04}$\\
ZTF18abecbks     & SN\,2018ddu  & SN Ic & 16:35:46.53 & +71:41:15.1 & 0.030 & 0.12 & 16:35:45.74 & +71:41:12.5 & $-20.23^{+0.46}_{-0.05}$\\
ZTF20aaiftgi     & SN\,2020aut  & SN Ic & 14:10:13.35 & -06:49:20.7 & 0.034 & 0.10 & 14:10:14.29 & -06:49:20.6 & $-21.53^{+0.07}_{-0.05}$\\
ZTF21aannoix     & SN\,2021dwg  & SN Ic & 14:18:15.96 & +00:53:18.4 & 0.026 & 0.11 & 14:18:14.91 & +00:53:27.9 & $-21.02^{+0.25}_{-0.04}$\\
ZTF21aaufwyh     & SN\,2021jao  & SN Ib & 10:20:52.91 & +06:09:24.1 & 0.028 & 0.06 & 10:20:53.36 & +06:09:22.3 & $-19.88^{+0.25}_{-0.06}$\\
ZTF19aakpcuw     & SN\,2019bgl  & SN Ic & 17:22:03.03 & +59:06:53.3 & 0.031 & 0.08 & 17:22:04.76 & +59:06:47.5 & $-20.75^{+0.09}_{-0.04}$\\
ZTF20aajcdad     & SN\,2020bcq  & SN Ib & 13:26:29.65 & +36:00:31.1 & 0.019 & 0.04 & 13:26:28.54 & +36:00:37.0 & $-20.75^{+0.29}_{-0.04}$\\
ZTF19acmelor     & SN\,2019uff  & SN Ib & 00:19:13.27 & -14:23:52.1 & 0.027 & 0.09 & 00:19:13.77 & -14:23:48.2 & $-20.44^{+0.05}_{-0.04}$\\
ZTF19abqmsbk     & SN\,2019orb  & SN Ic & 17:40:34.75 & +14:52:47.9 & 0.027 & 0.24 & 17:40:34.76 & +14:52:47.9 & $-18.81^{+0.13}_{-0.07}$\\
ZTF20aalcyih     & SN\,2020bpf  & SN Ib & 06:55:23.49 & +27:43:19.0 & 0.018 & 0.20 & 06:55:23.52 & +27:43:18.0 & $-18.46^{+0.22}_{-0.04}$\\
ZTF20abaszgh     & SN\,2020ksa  & SN Ib & 10:59:27.94 & +46:07:28.4 & 0.022 & 0.05 & 10:59:27.35 & +46:07:20.5 & $-21.81^{+0.05}_{-0.04}$\\
\enddata
\tablecomments{The host galaxy photometry was measured from SDSS and PanSTARRS images using 
{\tt LAMBDAR} \citep{Wright2016a} and translated to absolute magnitudes by modelling the host-galaxy spectral energy distributions with 
{\tt Prospector} \citep{Johnson2021a} following \citet{Schulze2021a}. The absolute magnitudes are corrected for MW extinction but not for host attenuation. The coordinates of the SNe and their host galaxies are measured in the J2000 reference system. About $2''$ from the center of the host galaxy of SN 2019ieh is another object. It is uncertain whether this is a star-forming region or a separate galaxy.}
\end{deluxetable}

\begin{deluxetable}{lcccccccccc}
\tablewidth{0pt}
\tabletypesize{\scriptsize}
\tablecaption{Supernova light curve properties \label{table:SNLCproperties}}
\tablehead{
\colhead{ZTFID} &
\colhead{t$_{0}$} &
\colhead{M$^{\rm{peak}}_{r}$} &
\colhead{M$^{\rm{peak}}_{g}$} &
\colhead{(g-r)$_{10}$} &
\colhead{$\tau^{\rm{rise}}_{r}$} &
\colhead{$\tau^{\rm{rise}}_{g}$} &
\colhead{$\tau^{\rm{fall}}_{r}$} &
\colhead{$\tau^{\rm{fall}}_{g}$} & 
\colhead{t$^{\rm{rise}}$} \\
\colhead{} &
\colhead{(jd)} &
\colhead{(mag)} &
\colhead{(mag)} &
\colhead{(mag)} &
\colhead{(day)} &
\colhead{(day)} &
\colhead{(day)} &
\colhead{(day)} &
\colhead{(day)} 
}
\startdata
ZTF19abauylg & 2458672.53 & -19.39 (0.01) & -19.27 (0.02) & 0.67 (0.05) & 2.73 (0.03) & 2.24 (0.02) & 16.34 (0.13) & 9.86 (0.08) & 11.49 (-0.02, 0.02) \\
ZTF19abfiqjg & 2458686.51 & -19.38 (0.04) & -19.18 (0.04) & 0.71 (0.04) & 4.56 (0.27) & 4.10 (0.19) & 19.19 (1.30) & 9.94 (0.72) & 13.93 (-0.84, 0.13) \\
ZTF19abztknu & 2458770.89 & -19.04 (0.03) & -18.70 (0.05) & 0.72 (0.04) & 7.63 (0.18) & 5.34 (0.09) & 33.17 (0.90) & 12.84 (0.44) & 36.26 (-0.94, 1.93) \\
ZTF20acvebcu & 2459205.77 & -19.11 (0.03) & -18.75 (0.04) & 0.59 (0.07) & 6.19 (0.26) & 4.47 (0.14) & 25.52 (0.99) & 11.97 (0.73) & 23.37 (-1.84, 1.18) \\
ZTF18abecbks & 2458315.93 & -17.89 (0.02) & -17.43 (0.02) & 0.77 (0.06) & 3.99 (0.08) & 4.29 (0.10) & 25.82 (0.33) & 17.35 (0.40) & 13.07 (-0.29, 0.20) \\
ZTF20aaiftgi & 2458885.06 & -17.92 (0.04) & -17.71 (0.05) & 0.61 (0.08) & 5.03 (0.58) & 3.88 (0.27) & 21.33 (2.30) & 11.07 (1.10) & 18.63 (-3.58, 3.97) \\
ZTF21aannoix & 2459292.43 & -17.90 (0.02) & -17.49 (0.02) & 0.77 (0.07) & 2.11 (0.35) & 2.43 (0.16) & 27.92 (0.52) & 19.08 (0.53) & 20.63 (-0.49, 0.64) \\
ZTF21aaufwyh & 2459338.45 & -17.86 (0.02) & -17.63 (0.02) & 0.67 (0.04) & 5.05 (0.11) & 3.97 (0.07) & 16.80 (1.33) & 9.67 (0.55) & 25.07 (-0.59, 0.41) \\
ZTF19aakpcuw & 2458542.32 & -17.76 (0.02) & -17.40 (0.02) & 0.63 (0.13) & 2.29 (0.27) & 3.01 (0.71) & 32.11 (0.60) & 22.16 (0.48) & 16.54 (-4.36, 1.60) \\
ZTF20aajcdad & 2458887.76 & -17.60 (0.01) & -17.59 (0.01) & 0.68 (0.05) & 2.34 (0.03) & 2.12 (0.02) & 17.62 (0.18) & 9.52 (0.10) & 14.19 (-0.27, 0.23) \\
ZTF19acmelor & 2458802.23 & -17.50 (0.04) & -17.08 (0.06) & 0.72 (0.07) & 3.15 (0.43) & 3.78 (0.31) & 33.38 (1.63) & 17.25 (1.67) & 13.37 (-2.44, 0.98) \\
ZTF19abqmsbk & 2458733.76 & -17.59 (0.06) & -17.23 (0.04) & 0.70 (0.05) & 3.74 (0.35) & 4.41 (0.12) & 26.05 (1.39) & 13.09 (0.69) & 18.73 (-0.65, 0.67) \\
ZTF20aalcyih & 2458899.23 & -16.76 (0.02) & -16.48 (0.02) & 0.69 (0.09) & 3.58 (0.12) & 3.59 (0.12) & 36.75 (1.19) & 11.64 (0.97) & 23.08 (-1.13, 0.68) \\
ZTF20abaszgh & 2458997.73 & -16.73 (0.03) & -16.73 (0.02) & 0.56 (0.16) & 2.57 (0.20) & 1.76 (0.11) & 11.29 (0.69) & 9.96 (0.46) & 6.33 (-0.72, 0.25) \\
\enddata
\end{deluxetable}

\end{document}